# A Synthetic Macroscopic Magnetic Unipole

Emmanouil Markoulakis[1], John Chatzakis[1], Anthony Konstantaras[1] and Emmanuel Antonidakis[1]

[1] Department of Electronic Engineering, Computer Technology Informatics & Electronic Devices Laboratory, Hellenic Mediterranean University, Chania, Greece

E-mail: markoul@hmu.gr

## Abstract

We demonstrate experimentally with a prototype for the first time, that an artificial emergent magnetic unipole hedgehog field in a simply connected three-dimensional domain is possible, emulating effectively the Dirac model by simply using a novel permanent magnets' topology and geometrical arrangement in a ring array. Although similar effects were demonstrated by others over the last decade, with these effects usually observed and lasting for a small fraction in time and applied at the quantum or microscopic scale using primary BECs, Skyrmions, Spin Ice and recently Chiral Magnets this was never shown until now and performed at the macro scale and by using normal magnets. The synthetic magnetic unipole ring array prototype progressively twists and steers the magnetic flux into vanishing curl field (i.e. vortex) geometry towards the centre of the ring and its air cap. This mere act alone proves enough to create the desired effect and an apparent isolated magnetic unipole region is observed at the centre of this magnetic ring array, as we have mapped with a 3-axis magnetometer and show with the quantum magnetic-optic device flux viewer, the ferrolens. This apparent unipolar magnetic array exhibits anomalous behavior at its centre, air cap and registers with a magnetometer or a normal compass needle as having the same magnetic polarity on both opposite faces of the ring on the same axis passing through the centre of the ring. Also the other opposite polarity pole is dispersed and confined along the periphery of the ring array with a 90° angle to the unipole axis and none existing in the far field and nowhere to be found. Even in the near field, counting the total number of discrete poles of the magnets of the array including its emergent unipole at the centre, counts to an odd number. The results were analyzed and software simulated and some interesting conclusions are drawn about the ultimate quantum vortex nature of ferromagnetism.

## 1. Introduction

First of all we have to clarify here that we consider that the term "magnetic monopole" regardless of what adjective word you put in front, is strictly reserved to describe the elusive hypothesized to exist natural quantum elementary particle. Therefore in order to avoid confusion we prefer instead to use in our research the term "synthetic magnetic unipole (i.e. uni-one)" to refer to the artificial human made implementation attempt to physically, loosely emulate the theoretic Dirac model for magnetic monopoles which is a dipole where the two opposite magnetic charge monopoles are separated and connected by a semi-infinite line thus infinite long and infinite thin curve in space, the Dirac String. So even Dirac [1] tells us in his theory that a totally isolated magnetic monopole is not possible to exist in any way as Maxwell [2] does, therefore essentially and exclusively magnetism is a dipole phenomenon unless proven otherwise in the future. It is our understanding and belief that a completely isolated hypothesized magnetic monopole in the rare case it can exist will have nothing magnetic about it and therefore is an impossibility and the whole term in this specific interpretation of it, is a contradiction.

However, with that being said there are still potential significant scientific and technological merits today from theoretical and experimental research on this field and specially the last years from loosely synthetic implementations [3] of Dirac theorized monopoles usually artificially induced via an external magnetic field or laser steering on Bose-Einstein Condensates [4][5] but sometimes also natural occurring and observed when brought at very low temperatures as topological texture defects, thus skyrmions lattices in a condensed or soft matter medium like Spin-Ice materials [6][7] or recently observed and manipulated in chiral magnets [8–12]. All above described are created or existing at the quantum or microscopic size scale as non-isolated magnetic unipole quasi-particles [13] . Parenthetical, although a bit outside the context of our research presented herein but not entirely, as we will show later on, we must reference that it is theorized and with some indirect experimental reasoning that Dirac magnetic monopoles are hidden inside ferromagnetic materials and are correlated closely to the Anomalous Hall Effect (AHE) [14].

What in general these quasiparticles magnetic unipoles have in common is that they are topological defects emergent phenomena, in the form of an energy vortex in the medium that resembles as an effect and therefore emulates the theorized Dirac magnetic monopole. Also, their monopole field is not inherent from any applied external field but merely intrinsic due to the angular momentum of their vortex motion and in most cases these vortices are short lived. The vortex nature of the theorized magnetic monopoles is actually what Dirac geometrically is describing in his work. A Dirac monopole is a point-like source of a possibly artificial magnetic field that forms at the endpoint of a quantum whirlpool. In quantum mechanics, an electron is described as a diffuse wave-like object rather than a point-like particle. Paul Dirac was the first person who understood the importance of studying the end points of quantum-mechanical whirlpools within these electron waves. He noticed that when an electron has such a terminating vortex, a magnetic monopole inevitably forms at the end point (i.e. apex of the vortex). A terminating vortex is the defining characteristic of the Dirac monopole magnetic charge.

In our experiments we adopted and implemented the same theoretical approach but applied it at the macroscale by using normal neodymium permanent magnets in a novel ring array formation which progressively twists the magnetic flux of the individual magnets in the array to an end point at the center of the ring and air cap of the ring array and therefore creating an artificial magnetic flux whirlpool. This magnetic flux whirl opposite to other approaches for creating synthetic magnetic unipoles which are usually short lived, is permanent thus stable. Also the synthetic magnetic unipole field in our case is not created by the induced vortex motion topological defect in a medium caused by an external applied magnetic field or other method of steering but instead is intrinsic and originating from the Net Quantum Magnet Field (QMF) [15][16] existing inside our unipolar magnetic array material and then projected to the outside, at the center of the air cap of the ring array in a small region a couple of millimeters. This array of individual magnets acts like a magnetic flux twister focusing and vortexing the flux coming from all the like poles of its individual magnets facing to the same angular direction in the circle (i.e. clockwise or counterclockwise) to an end point, apex tip of the vortex located at the center of the hole of the ring array.

In other words, the net magnetic field of our array prototype contains itself the synthetic magnetic unipole at the tip, end point of its magnetic flux vortex field formation and no other medium is needed in order to generate the unipole by angular momentum.

Considering also what is theorized by Dirac about magnetic vortices end points and also the possibility of Dirac magnetic monopoles hiding inside ferromagnetic materials responsible to one extend for the Anomalous Hall Effect as concluded previously in the literature [14] and inferred also by some experimental deduction, there is a fair chance our prototype array to act as a Dirac magnetic monopole trap in the best case scenario or inducing a magnetic unipole quasiparticle in the worst case.

We did not use any static magnetic shielding on our prototype and it is not to be confused with the well known Halbach magnetic array configurations [17][18] which suppress the one net pole of the array (i.e. weak pole but you can still measure a field) relative to the other net pole polarity (i.e. strong pole) by using spatial partial field cancelation technique in one side of the array and by forcing unlike poles magnetization axes in close proximity together whereas our synthetic magnetic unipole prototype ring array is doing something entirely different. It is fussing the magnetic flux coming from all the same sides of the discrete magnets in the array to the center of the ring and diffusing and spreading out along the perimeter of the ring, the flux coming from all the other pole sides of the magnets. End result in our prototype, both faces of the ring array have uniaxially the same magnetic polarity and along the perimeter of the array no field is present or can be measured for the far field (i.e. for distances larger than the diameter of the ring array).

Next, we will describe the materials, devices and methods we used in our experiments with the unipolar prototype and also its design and implementation details.

## 2. Materials and Methods

### 2.1 Ferrolens

As with our previous research we use the quantum magnetic-optic flux viewer ferrolens to probe inside the ferromagnets material and projects to the outside their net Quantum Magnet Field (QMF) [15][16][19]. This superparamagnetic $25\mu m$ thin film encapsulated inside a flat optical lens and physical device, displays a real time 2D imprint with some holographic depth, of the vorticity and curl of the magnetic flux inside the permanent magnet's bulk material.

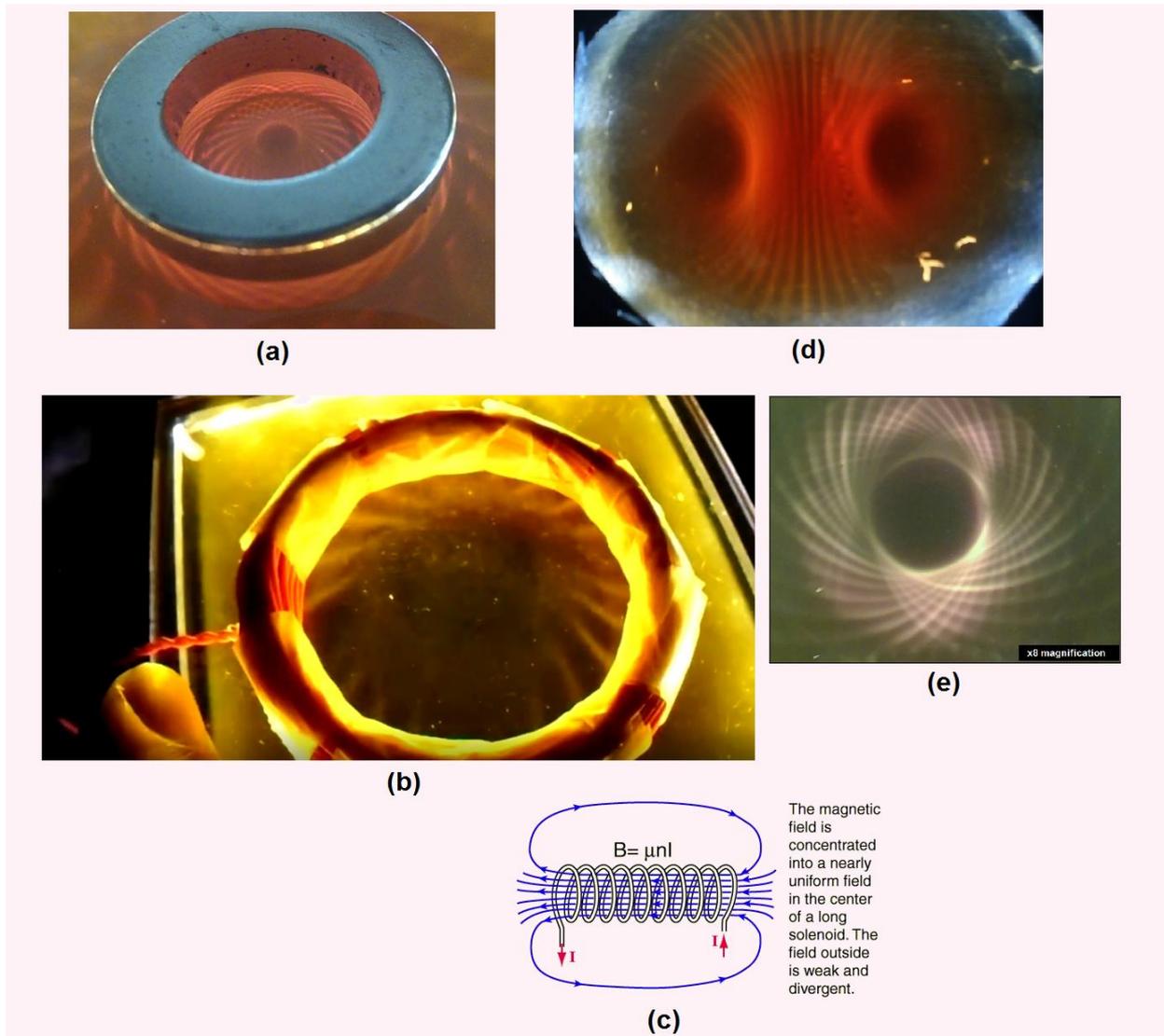

**Fig. 1** **(a)** Polar view of the field of a neodymium typical ring magnet shown on the ferrolens. The ring magnet has an axial magnetization by thickness (i.e. one face of the ring is the North Pole and the other the South Pole). Ferrolens shows the vorticity and curl of the magnetic flux inside the magnet's bulk material and projects it to the outside. The field displayed is confined inside the ring air cap. **(b)&(c)** The homogeneous field inside an electrical d.c. solenoid shown in real time by the ferrolens. Magnetic flux is going in straight parallel lines through the solenoid's length from one end (pole) of the solenoid to the other as theory predicts and also shown here by the ferrolens. The see-through property of the lens is demonstrated with holographic depth thus looking the one end of the solenoid we can see the flux lines terminating at the other end. **(d)** Neodymium cylindrical magnet put on its side under the ferrolens. Its two net Quantum Magnet Field (QMF) polar vortices are shown. Black holes left and right are the positions where the North and South Pole centers are physically located on the magnet. **(e)** Close-up magnified view of the vortex field inside the pole of a neodymium cube magnet.

In fig. 1(a) the polar view of the field of a typical neodymium ring magnet is shown by the ferrolens. The vorticity and curl of its magnetic flux lines is displayed in its air cap. The field shown depicts see-through both polar fields of the magnet (i.e. N and S) therefore the total field of the ring magnet is shown which is torus shaped. This is a unique property of the ring magnets when displayed with the ferrolens since they are able to confine the total field inside their air capped hole as shown by the lens. Fig. 1 (b) shows the inner magnetic field of an air capped without a ferromagnetic core, current solenoid, of which its theoretical model for its magnetic field is also illustrated in fig.1 (c). The inner field of an air capped electric solenoid in contrast to the field of a ferromanget, has no vorticity and nearly zero curl on its magnetic flux lines and is very homogeneous. This is nicely shown by the ferrolens in this particular snapshot, fig.1(b), from the experiments. You are looking in this snapshot, see-through the field on both magnetic poles N and S of the solenoid located at its two ends. The magnetic flux lines closer to the inner walls of the solenoid have larger curl than the lines at its center region which are straight with zero curl and vertical to the ferrolens surface and cannot therefore scatter light thus they are not made visible on the lens. The lines however close to the inner surface of the solenoid which are more curled are nicely picked up by the ferrolens and displayed with holographic depth.

## 2.2 The Magnetic Unipolar Array

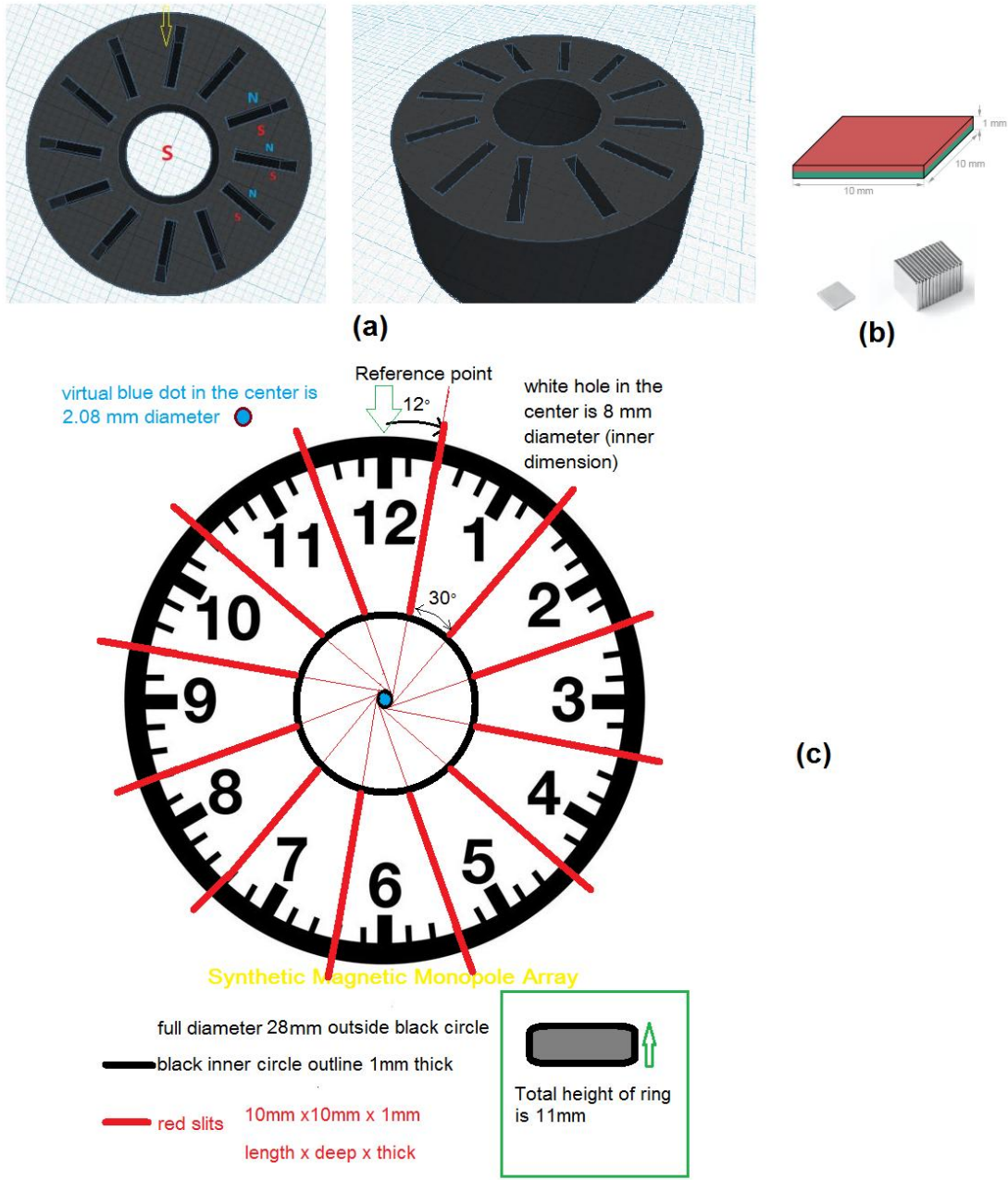

**Fig. 2 (a)** Illustrations of the Top and Top-side views of the 3D printed unipolar magnetic array's prototype v3, frame. **(b)** N42 neodymium magnetic plates used in the implementation of the unipolar array. Twelve magnetic plates are inserted vertically each into the corresponding frame slots shown. The polarity of the individual magnets inside the array follows a heads to tails pattern N-S-N-S... etc around the ring. **(c)** Drawing of the unipolar array's final prototype design v4. The type North or South of the emergent unipole formed inside the blue circle area, depends on the angular direction of the of the magnetic moments of the individual twelve magnets in the ring array, clockwise cw or counterclockwise ccw in combination with the 12° skew angle direction of each magnet plate in relation to the concentric radius line each time. Thus, if the magnetic plates at the ring array's perimeter lean to the left or to the right.

The magnetic array's frame shown in fig.2 (a) is designed in such a way that the axis of each magnetic N42 plate (see fig. 2b) when inserted vertically into a slot is not exact concentric to the circle's center as shown in fig. 2(c) and therefore not aligned with the imaginary radius line but instead intentionally eccentrically skewed 12° so that each magnetic plate axis does not point the ring's exact center but is instead tangent to a virtual 2.08mm calculated circle shown with blue in fig. 2(c). We have to stress out here how much important it is that the magnetic plates used in the design of the unipolar array to be as thin as possible else the magnetic flux of the array will not be steered and twisted precisely at the center of the ring array and the desired unipole effect will not be achieved. The angular separation of the twelve magnets used in the design is 30°. Each magnetic plate has 10mmx10mmx1mm dimensions and is a N42 grade neodymium axial magnetized magnet.

Type N or S, of the emergent unipole at the center of the ring as also stated at the legend of fig. 2 depends on angular direction of the magnetic moments of the magnetic plates, clockwise (CW) or counterclockwise (CCW), cascaded in series around the ring and in combination with their skew 12° angle direction left or right. The various combinations and the emergent each time unipole type is shown below:

**Table. 1** Unipolar Array Combinations

| Magnetic moment angular direction | Leaning of magnetic plates | Resulting unipole |
|---|---|---|
| CW | Right | *N* |
| CW | Left | *S* |
| CCW | Right | *S* |
| CCW | Left | *N* |

As a rule of thumb to the above Table. 1, if the leaning of the magnetic plates is in the same direction as the angular direction of the magnetic moments of the individual magnets around the ring then the emergent unipole at the center of the ring is always a *North (N)* unipole. Likewise, when leaning is at opposite direction of angular momentum then this results always to a *South (S)* unipole. As an example the angular magnetic moment in fig.2(a) goes CCW around the ring and the magnet plates are leaning to the right therefore the resulting unipole at the center is *S*.

Actually we have constructed in total four prototypes with prototype v4 being the final, we have used in our experiments. The slots for the magnetic plates shown in fig.2(a) of the prototype v3 we 3D printed, were actually in version 4 open ended on the perimeter of the ring for an easier installation of the magnetic plates inside the frame. This is shown in fig. 2(c) of the design drawing of prototype version 4 with magnet plates represented with the red thick lines shown. Also, the dimensions of prototype 3 have been

shrunken in prototype 4 for a more compact design. The 3D

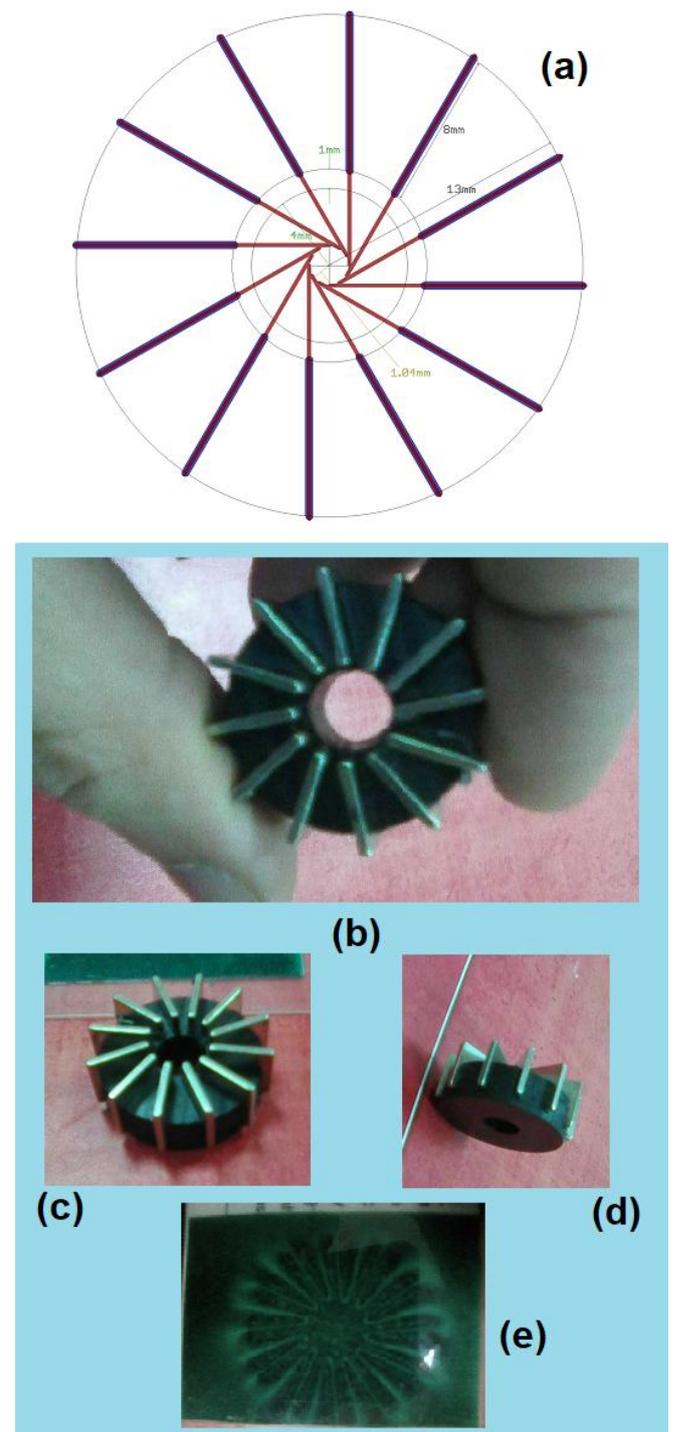

**Fig. 3 (a)** CNC Blue print of final prototype 4. **(b)&(c)** Top (Active side) and top-side views of prototype version 4. **(d)** Backside of unipolar array prototype. **(e)** Field 2D imprint of prototype 4 front active side, shown with a magnetic field viewing film.

printer files for prototype version 3 can be found in this link, https://tinyurl.com/ycw9uf5p. Final *prototype version 4* was not 3D printed but CNC machined from hard PVC material for extra stiffness. The CNC blue print and final constructed prototype four are shown in fig. 3. A very large size singled out image of the prototype's v4 CNC Blue print shown in fig.3(a) can be found in this link, https://tinyurl.com/ybmxk2qf for anyone who is interested to replicate the design.

Fig. 3(b)(c)(d), show different sides of the CNC machined prototype v4. Specifically, fig.3(b)&(c) show the front side of the unipolar ring array finished prototype PVC frame with the assembled in 12 magnetic plates. We call this front side of the prototype from now on as Active Side to distinguish it more rigorously from its backside shown in fig. 3(d).

Importantly, notice here that there is actually no difference except a small difference in field strength we will discuss later on, between the field at the front side Active Side of the unipolar array and its backside. Both sides of the prototype four register as a South type *S unipole*. This is because according to Table.1, when the ring array is viewed from its front, the so called active side, the angular magnetic moment of the ring array turns clockwise around (CW) with the magnet plates leaning to the left therefore an *S unipole* at the center of the ring. Similarly, when the ring array is viewed from its backside the situation of Table.1 reverses thus a counterclockwise (CCW) rotation of the magnetic moment with the magnets now leaning to the right therefore the end result as shown in Table.1 is the same as previously thus a *S unipole*.

In fig. 3(e), interestingly we see a 2D imprint of the resulting magnetic field geometry of the array using a commercial magnetic field viewing film. We see a teeth like structure is formed like arrows pointing to the center air capped hole of the ring array. We can count here the discrete 24 poles of the array on the perimeter and also observe the emergent unipole vortex forming at the center of the hole similar to a screw and apparently isolated from the rest of the field (see here a singled out larger size version, https://tinyurl.com/ybtof269). This **vortex unipole magnetization** is also shown in a close-up here in fig. 4.

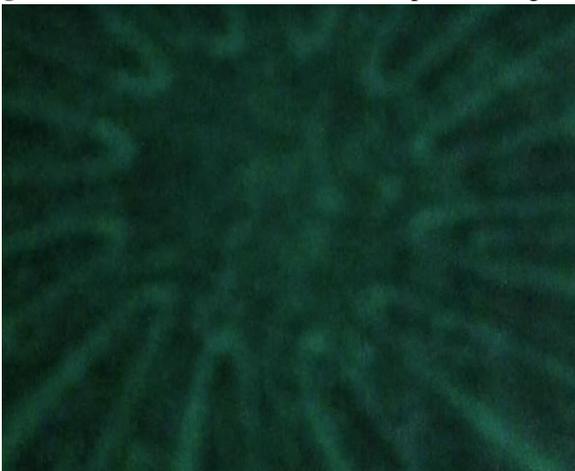

**Fig. 4** Screw like formation of the unipole vortex at the centre as shown with a magnetic field viewing film.

Notice in fig. 5 that a same circular magnetization array but without skewed angles on the magnets position introduced and instead all equatorial axes of the magnets being entirely concentric to the centre of the ring will produce no magnetic field at all at the air capped hole region of the ring or elsewhere outside the ring bulk material. Its field is similar to a toroidal solenoid where no field exists outside the solenoid.

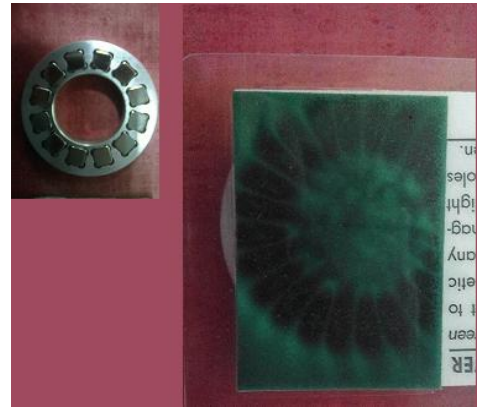

**Fig. 5** Circular aluminum frame concentric magnetization ring array. No field (i.e. light green color) exists outside the ring bulk material.

### 2.3 Other Instrumentation

A sensitive ±70µT range, 3-axis magnetometer with resolution 143mV/µT was used in conjunction with a digital oscilloscope for making the field strength measurements for the surface mapping of the field of the Unipolar Array prototype and by using also a xy-coordinates digitizer.

## 3. Results and Discussion

### 3.1 Decay of the Unipolar Array Field

It is well known that any point-charge electric or hypothetically magnetic, which has a spherical symmetry, would decay with distance from the source with the inverse square law $1/r^2$ contrary to dipole fields, of two opposite polarity separated charges which decay with the inverse cube law $1/r^3$. For non-point charges the above described are still true and hold for distances r>>D (*i.e. far field*) where D is the dimension of the non-point charge for monopolar fields or the line distance between the centers of the two charges for dipole fields. Also, the field of a hypothetical magnetic charge monopole would have a non-zero divergence and zero curl magnetic flux lines which would extent radially out in space thus the hedgehog effect.

Measuring the decay of the field strength, magnetic flux density B with distance on the z-axis, of our prototype unipolar array on the far field to see if it complies more to the inverse square law which is true for monopoles than the inverse cube which holds for dipoles, would possible provide important data about the actual nature of our prototype under testing and its characterization.

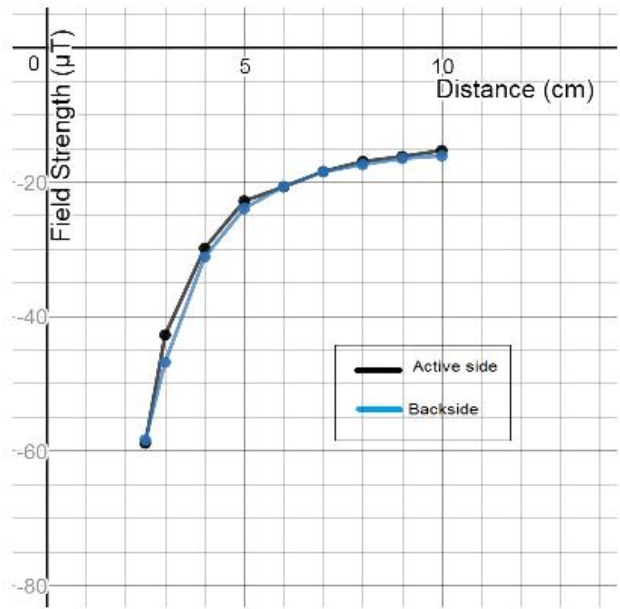

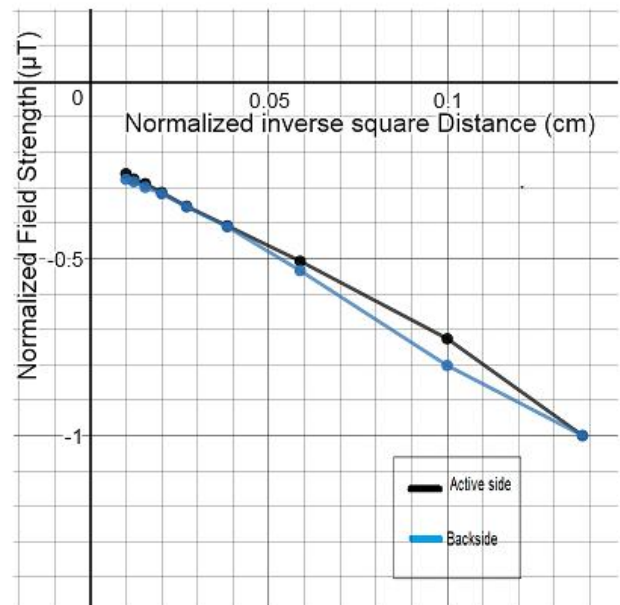

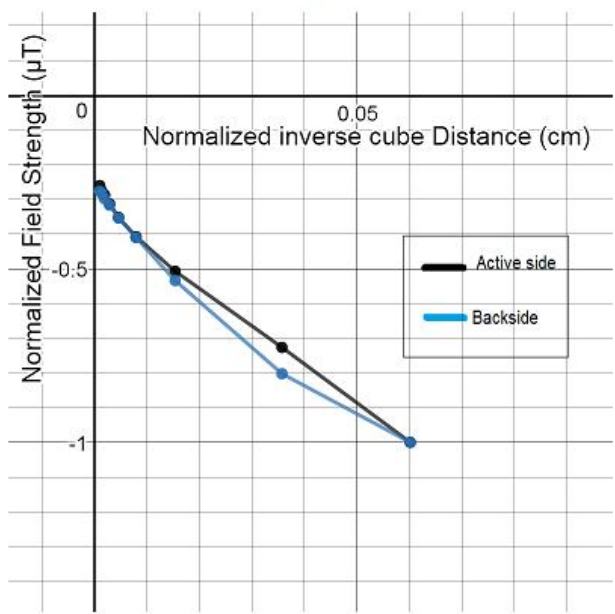

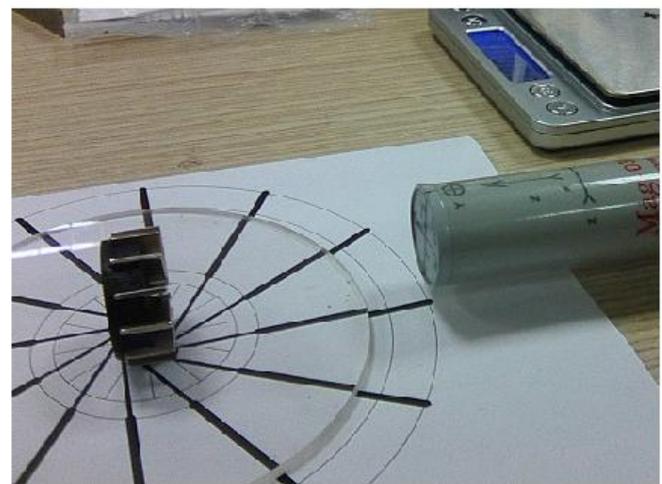

**Fig. 6 (a) (b) &(c)** Plots of z-axis field strength decay with distance measurements of the unipolar array prototype v4. **(d)** The experimental setup with the prototype and the 3-axis magnetometer. The magnetometer here is measuring the field only on the z-axis.

We see in fig.6(a) https://tinyurl.com/y7kb7kjc the linear xy axes plot of the decay of the prototype's z-axis field with distance. Fig.6(b) https://tinyurl.com/yde5bf42 shows the normalized values version plot to the inverse square distance. Ideally if the unipolar prototype complies with the inverse square law with distance, the plot will be a straight line. Similar fig.6(c) https://tinyurl.com/y7hcp2cl shows the normalized values to the inverse cube with distance. Both plots of the front, active side of the array and its backside (blue plot) are shown. Fig.6(d) shows the experimental setup. The z-axis position in space was chosen in such a way that the magnetometer does little as possible register the magnetic N-S vector field of the Earth although a total control experiment is very difficult because the curl of the magnetic flux of Earth's magnetic field and it would require heavy magnetostatic shielding in a zero-Gauss chamber.

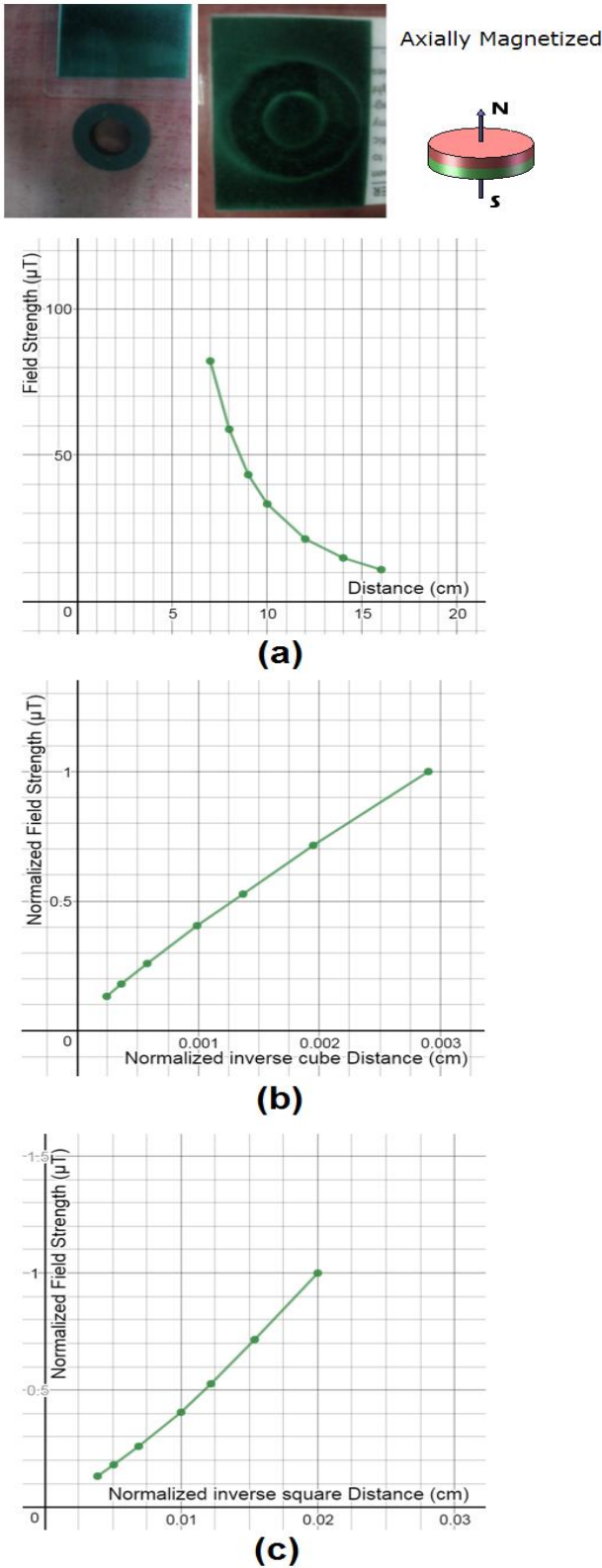

**Fig. 7** **(a)** **(b)** **&(c)** Field decay with distance of normal axial magnetized ring magnet and its normalized values plots.

We see in fig.6(b) that the far field measured values for the unipolar prototype draw in a straight line and therefore comply with the inverse square law. Only at the last most distanced measurement point thus at 10cm the values started to deviate from the inverse square law. This deviation of the normalized value of this last plot point can be contributed to the fact that at this point the measured value of the field strength of the array dropped at 17µT (see fig.6a) which is actually much smaller value than the Earth's magnetic field (i.e. 40 to 50µT) and the curl of the Earth's magnetic flux at that point started to significantly interfere with the measurements. Observing the plot of the normalized values to the inverse cube law with distance of the far field at fig.6(c), we see that in contrast to fig.6(b) of the normalized inverse square law plot, the values here start to deviate early on from a straight line and continue to do so as a whole therefore the measured field strength values of the unipolar prototype do not decay with the inverse cube law therefore inferring to the anomalous characteristics of our prototype's field.

As a reference for comparison we repeated the experiment for a normal axial magnetized ring magnet and draw the corresponding plots for the measured field decay values as shown in fig.7. We used a ferrite magnet which is much less strong than neodymium magnets so that we get comparable field strength values at the same distances, see fig.7(a) https://tinyurl.com/yayggfrb . As expected we see in fig.7(b) https://tinyurl.com/yaqnh4ks of the normalized inverse cube values, an almost perfectly straight line, therefore a normal dipole magnet's field decay complies with the inverse cube law with distance as measured thus the field decays exponentially with distance. Furthermore, in fig.7(c) https://tinyurl.com/yaz7m86p of the inverse square normalized plot, we see that the values deviate clearly and curve out from a straight line and therefore do not comply with an inverse square with distance decay.

As a final observation here we cannot fail to notice in fig.6 the close matching up of the prototype's measured field values from its front side (i.e. active side) and backside which infers to the symmetry of the unipole field. Distance was always measured from the plane of the magnetic plates to the magnetometer sensor as shown in fig.6(d). The extra plastic frame material thickness existing at the backside of the array was also taken into consideration so it does not mess up the distance measurements.

Next, we will continue the measurements for the prototype's field characterization by graphing its magnetic field surface maps.

### 3.2 Surface mapping of the Unipolar Array's Field

The discrete measurements for the surface mapping of the field of the array were taken at 30° intervals around the ring faces both front and backside of the prototype, at the exact positions where the magnetic plates are located inside the

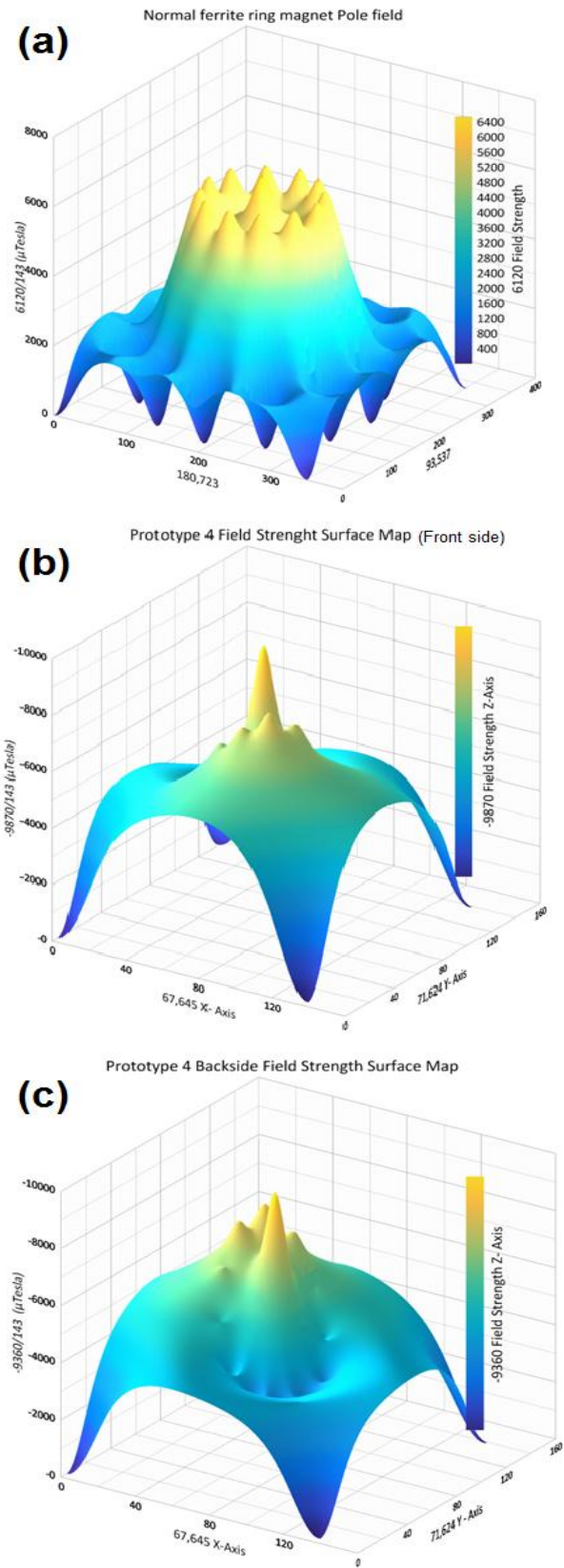

**Fig. 8** (a) Normal ring magnet. (b)&(c) Prototype Unipole.

prototype's frame as well as measurements at the exact center of the ring. For reference in fig.8(a) we mapped the field strength measurements of a normal axial magnetized ferrite ring magnet, namely its North Pole face. The values of the field strength on the z-axis of the maps are calculated in µT units by reading the output in mV of the oscilloscope (OSC) connected to the magnetometer and applying the equation (1), Distance from magnetometer sensor was 2cm.

$$B(\mu\text{T}) = \frac{OSC(mV)}{\text{Magnetometer Resolution}} \quad (1)$$

with the magnetometer's resolution given by the manufacturer at 143mV/µT. Negative values on the z-axis of the maps indicate a South Pole whereas positive values a North Pole. The formed yellow spikes on the rim of the map of fig.8(a) are due our discrete measurements points taken at regular intervals and the actual field of the ferrite ring magnet will have the same strength uniformly around the rim yellow region shown, with no spikes. We see however that the highest field strength at the far field of the ferrite ring magnet is measured at its exact center air capped hole.

In fig.8(b)&(c) we can see the surface mapped field of our unipolar ring array prototype v4 of its front side face and backside accordingly. By observing these two maps we quickly realize that it is the same unipole field observed by two opposite spatial regions although the unipole field emanating from the backside measures relatively at an average 2 µT less. Also, the field registers on both sides of the prototype axially as a *South S Unipole* with the highest field strength concentrated at the center air capped hole of the prototype ring array at the far field. Interestingly, we discover that the field around the frame of the unipolar array where the magnetic plates are located seems to dissipate progressively around the ring with a slope function which is a similar characteristic with the angular velocity function of an irrotational free vortex therefore strongly suggesting that a magnetic flux vortex is present in our synthetic unipole.

This will become more evident next were we map the field of the prototype ring array along its periphery at the far field, which is 90° perpendicular to its unipole axis, as shown in fig. 9 of the experimental setup. The distance from the magnetometer sensor is about 2 cm out to the periphery of the prototype and is hold fixed. The small humps and dips along the black dashed close loop line ellipsoid shown in fig.9 of the surface map are the measurement points and where the magnetic plates are located on the periphery of the prototype.

Something extraordinary is happening here emphasizing the anomalous characteristics of our unipole array prototype. As we can see, the field on the periphery at the first half 0° to 180° section of the ring array, registers as a South Pole, and progressively diminishes in strength from about -30µT (i.e. minus sign indicates a S Pole, -4400/143 see fig.9) to 0μT accordingly and then for the second half of the ring from 180° point, back to 0° it registers as a North Pole with an progressively increasing strength from 0μT to about +3μT (plus sign indicates a North Pole, 400/143 see fig.9). We notice that the ring array prototype has on half of its perimeter the same polarity S Pole as its unipole at the center of the ring and on the other half, a North Pole but greatly suppressed in strength relative to its South Pole on the periphery (see fig.9 illustration). As a net result the generated poles of our prototype including the emergent unipole at the center are three thus an odd number and therefore an anomalous result.

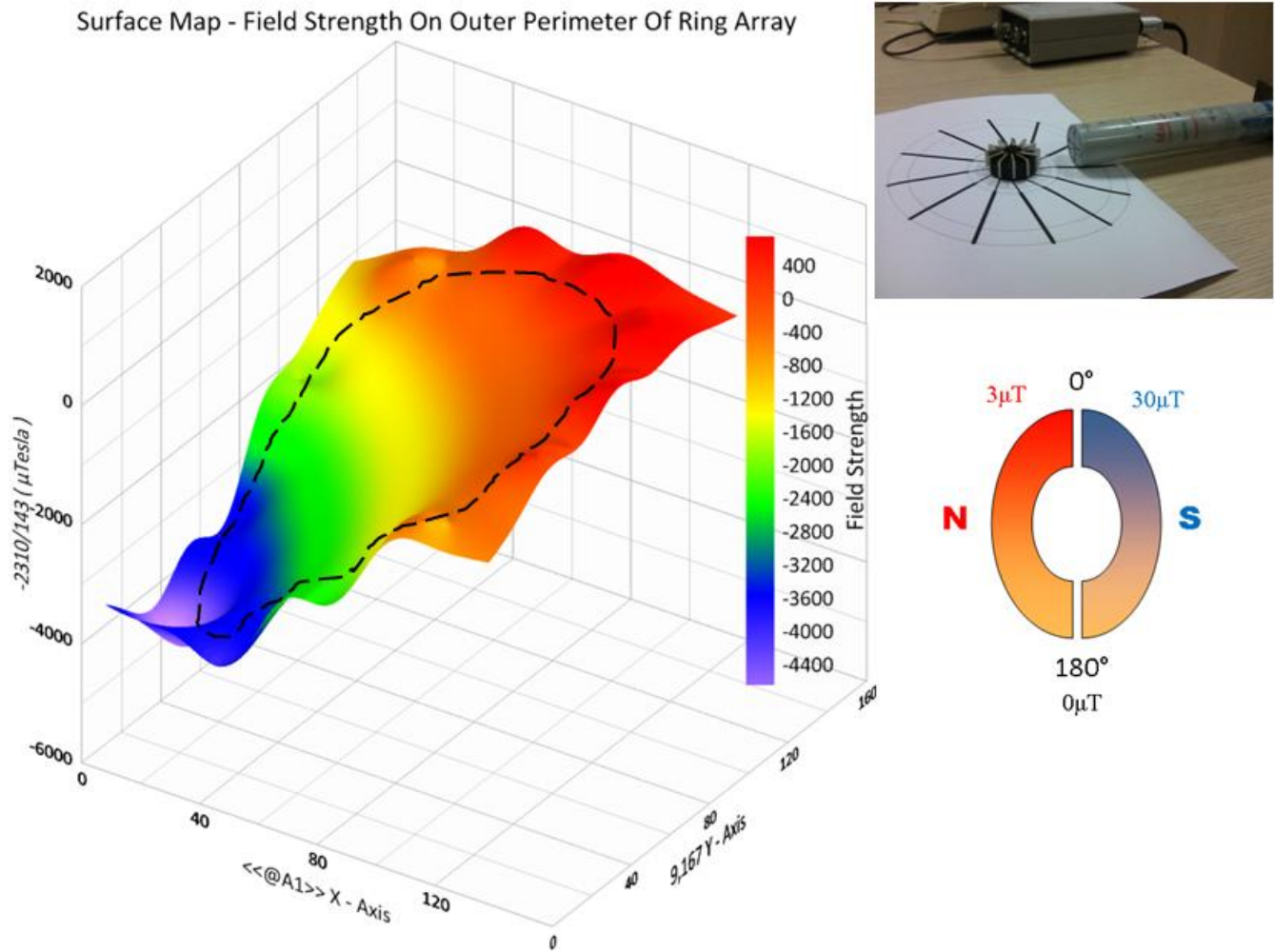

**Fig. 9** Surface map of the field on the periphery of the prototype unipolar array.

### 3.3 Ferrolens Observations of the Unipole Vortex

Although the prototype v4 of the magnetic unipolar array with a one axis Hall sensor directly upon the center of the front side or backside ring hole, at zero distance measures 22mT *S pole* field strength for the front and 17.6mT *S pole* for the backside of the prototype, (i.e. the difference between back and front is actually less than the one measured because the added 2mm plastic frame thickness on the backside of the ring array) and barely registers on a typical ferrolens which has a minimum sensitivity for an adequate display of the field at least 40mT, we actually succeeded on displaying in real time the formed apparent isolated induced unipolar vortex on the lens thin film from our prototype array. This was achieved by using the latest development generation of ferrolenses commercially available (i.e. Ferrocell commercial name) which have a more sensitive diluted thin film mixture and less thickness of the film (i.e. 509 type Ferrocell) [20].

In fig.10(a) we see the prototype placed on top of the ferrolens. Fig.10(b) is a black & white photograph for enhanced contrast, of the unipole magnetic flux vortex field of the prototype array as shown by the lens. The cross section of the black vortex formed at the center is no more than 3mm. This photograph was taken with the prototype placed under the ferrolens. Fig.10(c)&(d) show the vortex when viewed from the front side and backside of the prototype accordingly, through the prototype's ring array 8mm hole. In this case the prototype was placed on top of the ferrolens.

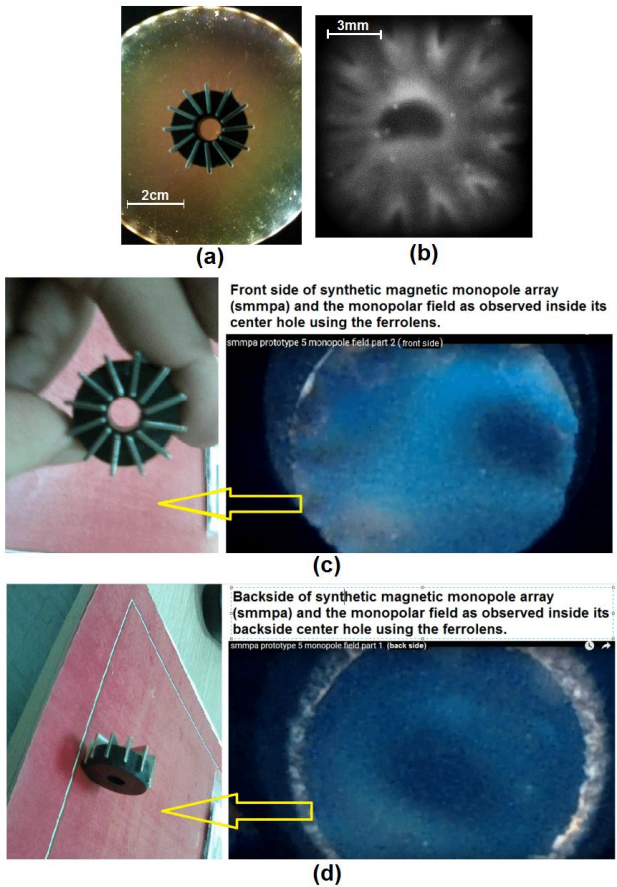

**Fig. 10** **(a)** Prototype unipolar array placed on top of the ferrolens. **(b)** Black and white photograph of the isolated unipolar magnetic vortex field of the prototype array as shown by the ferrolens. **(c)&(d)** color photographs of the unipolar magnetic flux vortex formed and displayed by the ferrolens. Photographs were taken through the 8mm diameter hole, front and backside of the prototype ring array's plastic frame.

This vortexing magnetic flux and apparent isolated vortex shown by the ferrolens is not the classical magnetic macroscopic field but the net Quantum Magnet Field (QMF) [15][16] inside the prototype ring array's material projected to the outside and shown by the ferrolens quantum magnetic optic physical sensor. It displays the vorticity and curl of the field inside the magnetic material of the prototype array which are different than the classical magnetic field formed outside the magnet on air.

If there is a unipole charge or net unipole charge, formed inside this vortex at its apex as predicted by Dirac theory [1], it would emanate magnetic flux radially with almost zero curl in the form of an hedgehog field to all directions and this would be the same for both QMF and classical magnetic field. Previously, in section 2.1 and fig.1(b) we have shown experimentally with the field inside an electric solenoid, that the ferrolens is capable of sensing and displaying straight without curl or very little, magnetic flux lines as long these are not perpendicular to the lens surface. It might be possible to see these straight flux lines with the ferrolens coming out radially from the vortex mouth to all directions as a hedgehog field, given a large relative optical zoom and a sensitive enough ferrolens. This is shown in fig.11.

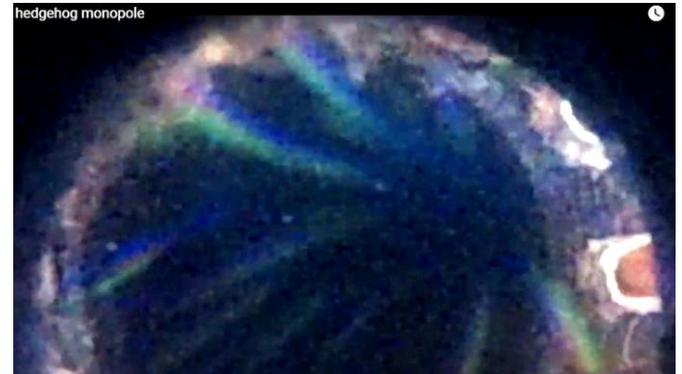

**Fig. 11** Hedgehog field lines displayed by the ferrolens, diverging out of the unipolar magnetic ring array's vortex field formed at the centre of the ring air capped hole.

Notice that the flux lines appear divergent coming from a center point, the vortex field.

Next we will simulate these results from our observations with the ferrolens of our Unipolar Magnetic Ring Array and from our analysis and data presented so far of all the experiments we have performed, thus a possible macroscopic synthetic magnetic unipole quasiparticle field device or even possible according the literature [14], a ferromagnetic Dirac magnetic monopole trap. Unfortunately, for direct observation of these particle charges, since the ferrolens observations cannot provide us with the submicron optical resolution and magnification needed, SQUID Scanning Magnetic Microscopy would be the best choice here, however we do not have this option yet although we are planning for it in the future. Therefore, we have to deduce the existence of these synthetic unipole charges indirectly by observing, measuring and analyzing their resulting field in our novel unipolar magnetic array.

### 3.4 Simulations of the Unipolar Magnetic Array

The simulations of the ferrolens display of the prototype unipolar magnetic array were executed with the commercially available software simulator program of Michael Snyder, pic2mag v1.0 Multithreading. It is the only scientific program currently as far as we know, that can simulate successfully the ferrolens induced magnetic field of magnetic arrays and its display. We will use pic2mag s/w to further verify our experimental results and observations and also see details of the total field of the prototype unipolar array which were not possible previously entirely to observe with the actual physical device ferrolens because the sensitivity limitations of the device we have described in section *3.3*. The near field, at the bulk magnetic material on the face and periphery of the ring array which is greatly confined inside the ring as we have measured and will show later on with more detail in the simulations, even at distances no more than 5mm falls off quickly at the z-axis to values varying from 7mT as we have measured with a Hall sensor to as low as a few µT and therefore not adequate field strength prototype. How the simulation works, you have to input first the exact magnetic array's geometry, dimensions. We have used a xy coordinate's digitizer to input the array's topology into the simulator. Then using RGB color coding on each discrete magnetic plate we inputted the exact spatial orientation of the magnetic moment of each individual magnet plate with reference to a 360° circle. The program then executes and simulates the ferrolens display of the field for displaying the magnetic flux in the array by the actual ferrolens. When a magnet is placed in contact with the ferrolens there is always a separation distance from the encapsulated thin film inside the optical lens to the magnet. This is because the thickness of the optical glass used by the lens. In our case we used ferrolenses with an optical glass thickness in between the magnet and the lenses' thin film of about 2.5mm.

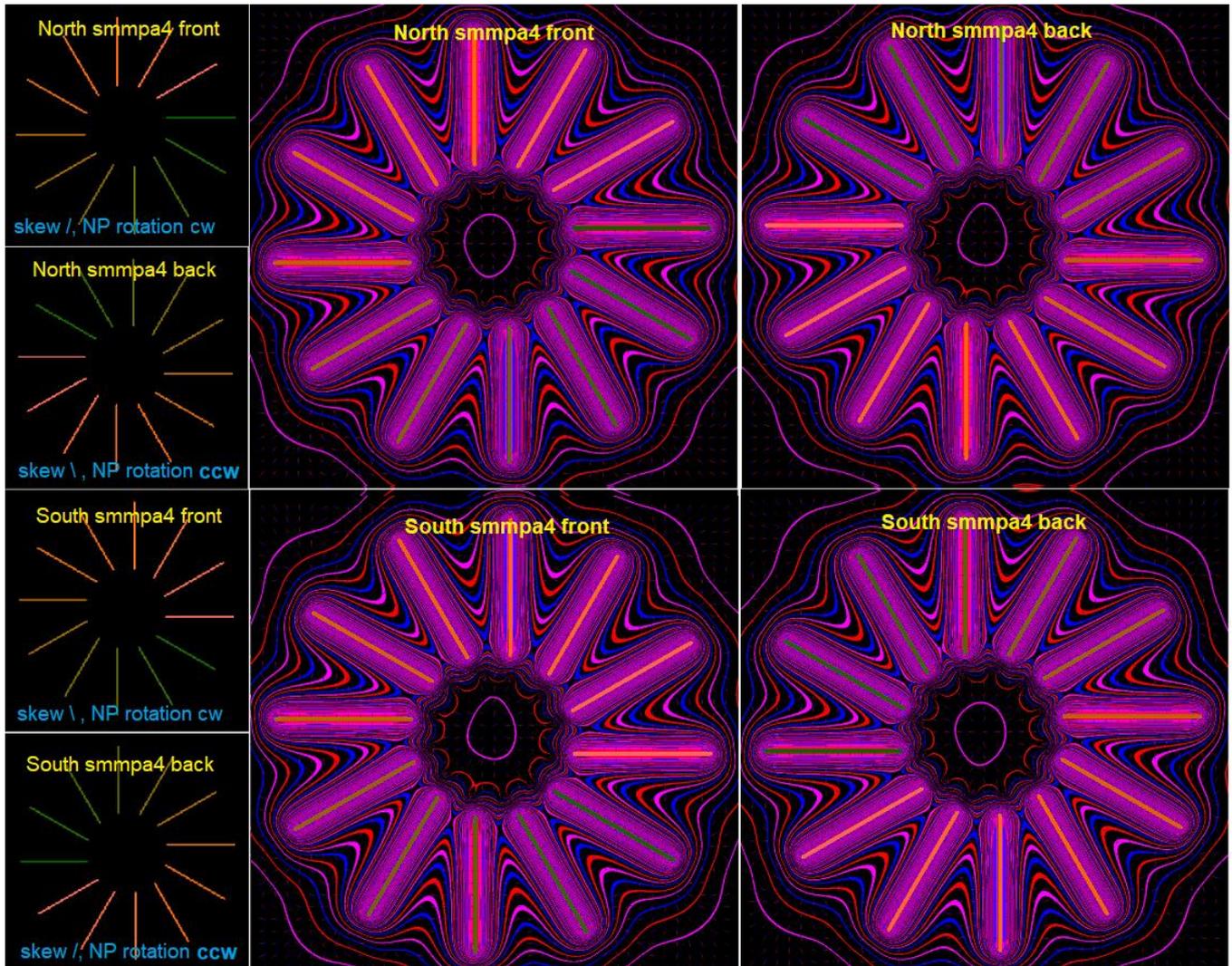

**Fig. 12** Simulations of the prototype v4 unipolar magnetic array in various configurations listed on Table.1. Both field views of the prototype unipolar ring array are shown when observed from its front side or backside.

The simulation has no such limitations and can display the magnetic flux vorticity and curl inside the magnetic array inside the prototype unipole array. In fig.12 we see the results of these simulations for various configurations of the prototype listed in Table.1 section 2.2 with the field viewed on both sides, front and back of the prototype unipole array. Thus, direction of skew angle of magnet plates left ( \ ) or right ( / ) and angular rotation direction of the magnetic moment of the ring array, clockwise (NP North Pole rotation CW) or counterclockwise (NP North Pole rotation CCW).

In all cases we see the vorticity and curl of the magnetic flux inside the array thus the Quantum Magnet Field (QMF)

of the unipolar array as displayed by the simulation. The software simulates the display of the ferrolens assuming a magnet's side is put under the lens and in contact. We clearly can see in all the simulations the emergent isolated unipole quantum magnet flux vortex displayed as an oval, egg shaped purple loop at the center of each array air capped hole. This actually is more or less what we have observed with the actual, physical ferrolens device and also measured in the other experiments with the magnetometers. Apart of the QMF of the prototype shown by the simulator, the simulator can also calculate the classical macroscopic on air, N-S vector magnetic field of the unipole array shown in fig.13.

The software simulator pic2mag v1.0 for illustrating the vector field uses small bicolored magnetic rods which act as tiny compass needles shown in fig.13. Unconventionally, it uses red for the South Pole of the rods and blue indicating their North Pole. In fig.13(a) we see a close-up zoomed in version of our South *S* unipole simulation at the center of the array's air hole. We observe that the vector field diverges to all direction from the center of the hole where the vortex is formed as Dirac theory predicts like a hedgehog. Also, the rods North poles (blue) point inwards therefore a South Unipole at the center of the prototype ring array. Fig.13(b) shows the total classical vector field.

### 3.5 Detection of magnetic unipole with a solenoid

Apart of using SQUID scanning magnetic microscopy for submicron mapping the synthetic unipole field at the center of the magnetic flux vortex, located at the center of the prototype ring array hole, see previously fig10,

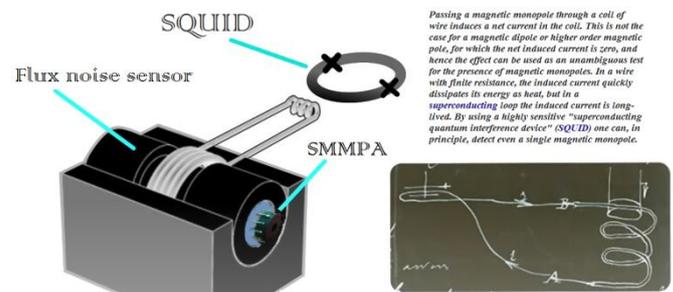

we can use also a SQUID magnetometer to simply detect the presence of magnetic monopole(s):

"Passing a magnetic monopole through a coil of wire induces a net current in the coil. This is not the case for a magnetic dipole or higher order magnetic pole, for which the net induced current is zero, and hence the effect can be used as an unambiguous test for the presence of magnetic monopoles. In a wire with finite resistance, the induced current quickly dissipates its energy as heat, but in a superconducting loop the induced current is long-lived. By using a highly sensitive "superconducting quantum interference device" (SQUID) one can, in principle, detect even a single magnetic monopole." [21].

The SQUID method of detection shown in fig.14 is very sensitive and most suitable for detection of magnetic monopoles or related quasiparticles at the microscopic scale like Spin-Ice materials and beyond and could theoretically even detect a single magnetic monopole inside matter. However, in our case of the synthetic magnetic mono (uni) pole prototype array (i.e. smmpa) and due its macroscopic ferromagnetic nature the above described experimental setup for detection of magnetic monopoles could be replicated but without necessarily the use of a SQUID magnetometer.

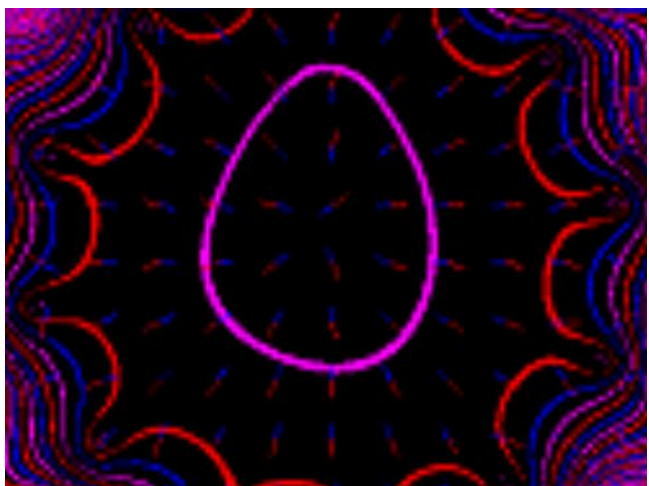

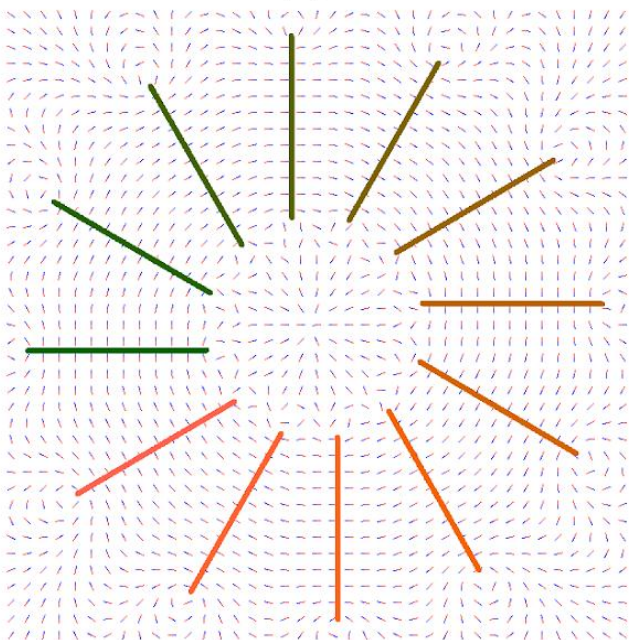

**Fig. 13** **(a)** Simulated vector hedgehog field of the *South* unipole. **(b)** Simulated vector field of the whole unipole array.

Instead, as shown in fig.15(a) a relative long electric solenoid was special made and wire coiled around a 3D printed plastic pipe and its two wire ends were connected with an R=1Ω resistor. The induced electric current waveform in the solenoid was displayed and measured with a digital oscilloscope connected across the shunt resistor. Various types of dipole ring magnets including our unipole smmpa array were allowed to free fall vertically using a wooden stick guide, to ensure that the magnets fall vertical with their pole and do not flip inside the solenoid, throughout the total length of the solenoid and the induced signal was recorded on the oscilloscope for comparison. The inductance of the solenoid used in the experiments was measured with a RLC bridge to be L=1.35 mH with an ohmic wire resistance of RL=10 Ω.

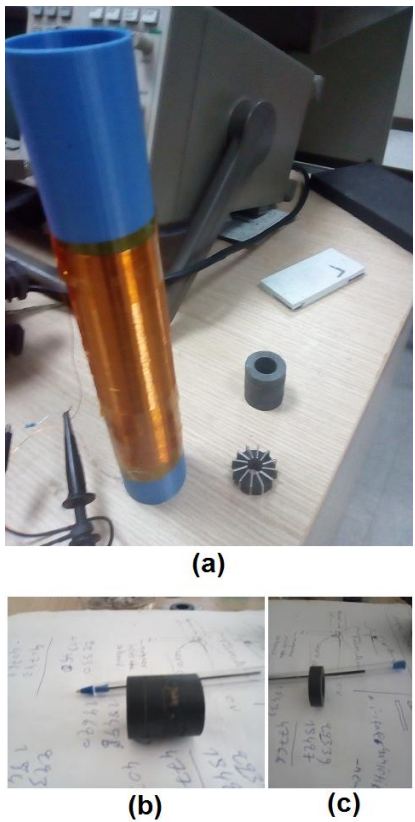

**Fig. 15 (a)** Experimental setup, a 15cm length solenoid about 750 turns of 0.2mm copper insulated wire coiled on a 25cm long 3D printed plastic pipe. Both wire ends of the solenoid are connected via a 1Ω shunt resistor and measured by a digital oscilloscope. The smmpa unipole ring array prototype and a large length ferrite dipole ring magnet are shown and used in the experiments free falling inside the solenoid vertically and the induced current signal on the solenoid was recorded. **(b)** Large length ferrite dipole magnet used in the experiments. **(c)** Short length ferrite dipole magnet used in the experiments.

In fig.16 we see the recorded signals correspondingly of the induced current inside the solenoid by dropping vertically the magnets inside the solenoid. The center segment of each waveform corresponds to the signal produced when the whole length of the magnet is inserted inside the solenoid and the magnet is free falling along the solenoid's total length. Surprisingly in fig.16(a), we observe that for the long ferrite dipole ring magnet of fig.15(b) when dropped inside the solenoid and as long its total length is inserted and travelling inside the solenoid, the induced current segment shown is not zero but a variable slope current and therefore does not initially comply with the predicted in theory [21] zero current in this case.

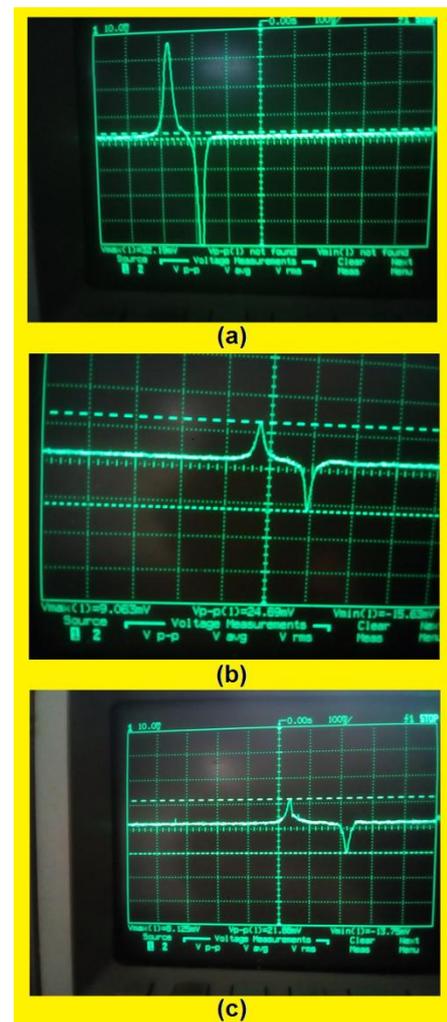

**Fig. 16** Recorded induced current signals from free fall dipole magnet experiments through a solenoid.

However, by further experimentation we discovered that this was due the fact that the length of the dipole magnet lm, was comparable to the length of the solenoid ls and therefore producing a ratio $l_s/l_m$ closer to one. In practice a much larger length solenoid or a much shorter dipole magnet will produce the desired effect and comply with the theoretical prediction. We decided to repeat the experiment with a much shorter dipole ferrite ring magnet shown in fig.15(c) and its induced current signal in the solenoid is shown in fig.16(b). As expected the net current segment at the center of the waveform shown in fig.16(b) is this time almost zero and fixed as theory ideally predicts validating therefore our previous observation and conclusion. The total waveform shown of the signal during entry and exit of the magnet from the solenoid appears asymmetric due the free fall gravitational acceleration of the magnet.

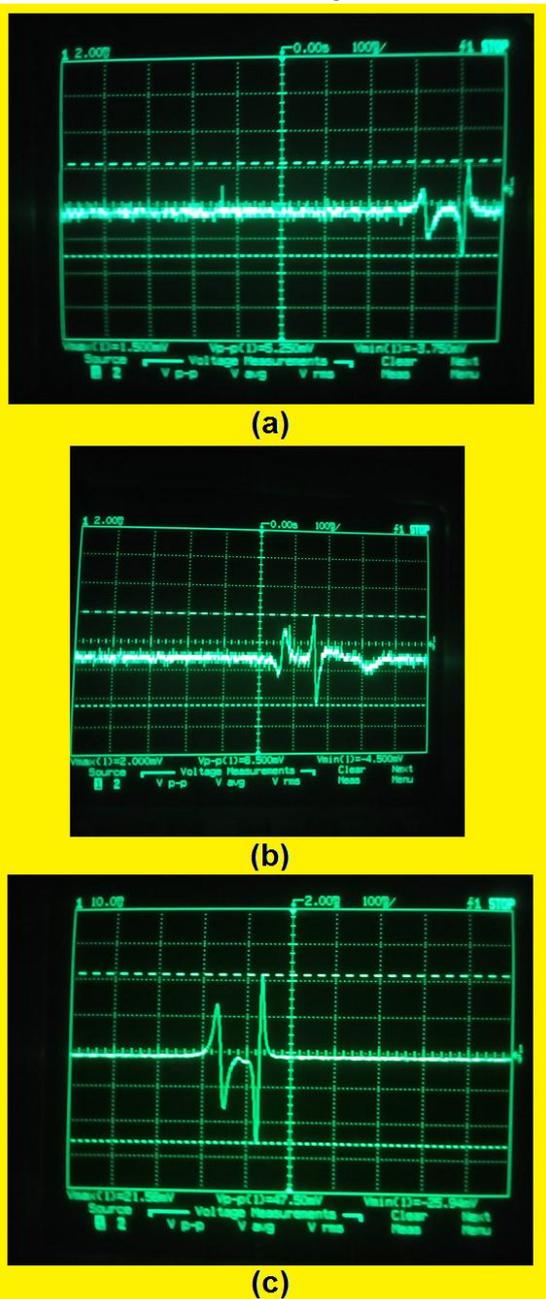

**Fig. 17** **(a)** South Unipole prototype solenoid current signal. **(b)** North Unipole prototype solenoid current signal. **(c)** South pseudo-unipole magnet solenoid current signal.

Additionally in fig.16(c), by applying magnetic braking we succeeded in extending the period where the magnet is inside the solenoid and therefore further demonstrating the net zero current flattening effect predicted by theory and also reducing the free fall acceleration therefore the total waveform appears more symmetric.

Next, we dropped our South Unipole prototype v4 ring array inside the solenoid at free fall. The produced induced current signal is shown in fig.17(a). It shows unambiguously that the middle current segment of the waveform where the whole length of the unipole is introduced inside the solenoid and traveling along its length, has a non-zero net current value. Therefore complying with the theoretical prediction about magnetic monopoles induced current in a solenoid [21]. Continuing, we dropped our smmpa prototype v3 in a North Unipole configuration through the solenoid and the recorded current signal is shown in fig.17(b). As expected the total signal waveform appears now inverted compared to the South Unipole signal shown in fig.17(a). As a side note, it is hard not to notice although irrelevant to our discussion here that the total waveform resembles strongly to the letter M. In all smmpa drop experiments the shunt resistor R of $1\Omega$ was replaced with a $10\Omega$ resistor value to increase the signal output measured by the oscilloscope since the measured magnetic strength of the prototypes smmpa are less than that of the typical dipole magnets we used in the experiments. Also, notice that the smmpa prototypes have about the same length as the short ferrite dipole ring magnet of fig.15(c).

Last but not least, we wanted to see if the length ratio of the solenoid over the unipole prototype length $l_s/l_{smmpa}$, affects in any way the generated current signal waveform. This is demonstrated at fig.17(c) where we constructed a relative large length pseudo-south unipole magnet by cutting the dipole ferrite magnet of fig.15(b) in half and forcing together by epoxy gluing the two formed North Pole ends at the middle forming therefore a single S-S pseudo unipole magnet. The recorded current waveform generated by dropping the pseudo-unipole magnet inside the solenoid is shown in fig.17(c). By comparing fig.16 with fig.17 waveforms we can see clearly, in contrast to dipole magnets that size of the unipole magnet has no effect on the general shape and symmetry of the induced current waveform, therefore further complying with the predictions deduced in theory.

## 4. Conclusions

By refining our research and observations [15][16][19] over the last years, we have come to the conclusion that ferromagnetism is a matter net dipole energy vortex macroscopic quantum phenomenon taking place inside the ferromagnet's material thus the net Quantum Magnet Field (QMF) which is different in curl and vorticity from the field outside the magnet on air, its classical macroscopic magnetic N-S vector field. With the net Quantum Magnet dipole vortex field inside the material being the cause and the classical field outside, the effect.

The primary drive for this research presented herein, is for answering the question what would happen if we artificially emulated what an actual Quantum Magnet is doing thus twisting magnetic flux energy into a free vortex whirlpool formation as predicted by Dirac P.A.M. in 1931. To our surprise, but not really, a magnetic pole was created out of the blue, validating our research about the ultimate intrinsic vortex irrotational nature of ferromagnetism, matter phenomenon.

We achieved our goal by using our research and experience in conceiving a novel ferromagnetic array and prototype introducing a unique as far as we know, vortex unipole magnetiztion which twists progressively the net Quantum Magnet flux into a vortex formation within the array's ring air hole at its center. Differentiating from most other research in the literature reported so far, apart that this is a macroscopic research and device for emulating synthetic Dirac unipoles at room temperature and not performed at the microscopic (e.g. Spin Ice) or quantum size level (e.g. BECs), in our case the emergent synthetic magnetic unipole at the center is permanent and is not due and created by the reaction response of another medium but intrinsic to the array itself and its projected Quantum Magnet field. **It is possible this to be the first** ever reported ferromagnetic Dirac monopole trap that isolates permanently a synthetic magnetic charge taken out within its own ferromagnetic matter. The existence of synthetic magnetic monopoles at room temperature inside ferromagnetic matter was theorized with some indirect experimental data, previously in the literature [14].

We have herein described in detail and characterized the device prototype, its magnetic field and operation, analyzed the results and experimental data numerically and by simulation and supplement related material [20]. For direct observation of the field we used the quantum magnetic optic flux viewer ferrolens we have used also in our previous research. We regret however that we could not have access for this research to SQUID scanning magnetic microscopy which has a submicron spatial resolution. We reserve this for future research and invite also other researchers to replicate our research. However, we have demonstrated by using experimentally an adapted simplified version of the solenoid induction current detection method [21] for magnetic monopoles in section 3.5 without the use of a SQUID magnetometer, that our macroscopic prototype synthetic magnetic unipole array unambiguously exhibits magnetic monopole behavior and therefore successfully emulates Dirac magnetic monopoles.

Although there are some other proposals and experiments done macroscopically to create or trap synthetic magnetic quasiparticles [22], our research as far as we know is the only presently pure magnetic by using normal permanent ferromagnets. The unique signal signature patterns of magnetic monopoles synthetic or natural, passing through a non-ideal ohmic solenoid we recorded for the first time and shown in fig.17 can be used by researches to aid them in the detection of magnetic monopoles.

## Acknowledgements

The authors are grateful to I. Rigakis, L. Frantzeskakis and Prof. E. Maravelakis for their assistance in the prototype versions construction and solenoid as well to B. Kerr, M. Snyder and T. Vanderelli for their support.

## References

[1]     Dirac P A M 1931 Quantised singularities in the electromagnetic field, *Proc. R. Soc. London. Ser. A, Contain. Pap. a Math. Phys. Character* **133** 60–72

[2]     Maxwell J C 1865 A dynamical theory of the electromagnetic field *Philos. Trans. R. Soc. London* **155** 459–512

[3]     Bender C M, DeKieviet M and Milton K A 2014 Synthetic versus Dirac monopoles arXiv:1408.4051

[4]     Ray M W, Ruokokoski E, Kandel S, Möttönen M and Hall D S 2014 Observation of Dirac monopoles in a synthetic magnetic field *Nature* **505** 657–60

[5]     Ollikainen T, Tiurev K, Blinova A, Lee W, Hall D S and Möttönen M 2016 Experimental realization of a Dirac monopole through the decay of an isolated monopole Physical Review X 7.2 (2017): 021023

[6]     Castelnovo C, Moessner R and Sondhi S L 2008 Magnetic monopoles in spin ice *Nature* **451** 42–5

[7]     Ladak S, Read D E, Perkins G K, Cohen L F and Branford W R 2010 Direct observation of magnetic monopole defects in an artificial spin-ice system *Nat. Phys.* **6** 359–63

[8]     Milde P, Kohler D, Seidel J, Eng L M, Bauer A, Chacon A, Kindervater J, Muhlbauer S, Pfleiderer C, Buhrandt S, Schutte C and Rosch A 2013 Unwinding of a Skyrmion Lattice by Magnetic Monopoles *Science (80-. ).* **340** 1076–80

[9]     Kish L 2018 *Skyrmions in Chiral Magnets* *https://tinyurl.com/yypg6up9*

[10]    Okumura S, Hayami S, Kato Y and Motome Y 2020 Tracing Monopoles and Anti-monopoles in a Magnetic Hedgehog Lattice (Physical Society of Japan)

[11]    Okumura S, Hayami S, Kato Y and Motome Y 2020 Magnetic hedgehog lattices in noncentrosymmetric metals *Phys. Rev. B* **101.14 (2020): 144416**

[12]    Kanazawa N, Nii Y, Zhang X X, Mishchenko A S, De Filippis G, Kagawa F, Iwasa Y, Nagaosa N and Tokura Y 2016 Critical phenomena of emergent magnetic monopoles in a chiral magnet *Nat. Commun.* **7.1 (2016): 1-7**

[13]    Pushkarov D I 2012 *Quasiparticle Theory Of Defects In Solids* (WORLD SCIENTIFIC)

[14]    Fang Z, Nagaosa N, Takahashi K S, Asamitsu A, Mathieu R, Ogasawara T, Yamada H, Kawasaki M, Tokura Y and Terakura K 2003 The anomalous Hall effect and magnetic monopoles in momentum space *Science (80-. ).* **302** 92–5

[15]    Markoulakis E, Konstantaras A, Chatzakis J, Iyer R and Antonidakis E 2019 Real time observation of a stationary magneton *Results Phys.* **15** 102793

[16]    Markoulakis E, Konstantaras A and Antonidakis E 2018 The quantum field of a magnet shown by a nanomagnetic ferrolens *J. Magn. Magn. Mater.* **466** 252–9

[17]    Halbach K 1980 Design of permanent multipole magnets with oriented rare earth cobalt material *Nucl. Instruments Methods* **169** 1–10

[18]    Halbach K 1985 Application of permanent magnets in accelerators and electron storage rings (invited) *J. Appl. Phys.* **57** 3605–8

[19]    Markoulakis E, Rigakis I, Chatzakis J, Konstantaras A and Antonidakis E 2018 Real time visualization of dynamic magnetic fields with a nanomagnetic ferrolens *J. Magn. Magn. Mater.* **451** 741–8

[20]    Google Drive - E. Markoulakis, HMU, 2020, https://tinyurl.com/y95xofdu

[21]    Glenn Cunningham 2019 High Energy Physics - Glenn Cunningham - Google Books *High Energy Physics* (Scientific e-Resources) p 264, https://tinyurl.com/ybjntsuk

[22]    Béché A, Van Boxem R, Van Tendeloo G and Verbeeck J 2013 Magnetic monopole field exposed by electrons *Nat. Phys.* **10** 26–9